\documentclass[aip,pra,twocolumn,showpacs]{revtex4}

\usepackage{graphicx,color}
\usepackage{amsmath}
\usepackage{amssymb}
\usepackage{longtable}

\begin{document}

\title{Accurate transport cross sections for the Lennard-Jones potential}

\author{S. A. Khrapak\footnote{Also with the Joint Institute for High Temperatures of the Russian Academy of Sciences, Moscow, Russia; \\ Electronic mail: Sergey.Khrapak@dlr.de}}
\date{\today}
\affiliation{Forschungsgruppe Komplexe Plasmen, Deutsches Zentrum f\"{u}r Luft- und Raumfahrt,
Oberpfaffenhofen, Germany}

\begin{abstract}
Physically motivated expressions for the transport cross sections describing classical scattering in the Lennard-Jones potential are proposed.
These expressions, which agree with the numerical results better than to within $\pm 1\%$, can be easy implemented in practical situations. Some relevant examples are provided.
\end{abstract}

\maketitle

\section{Introduction}
The problem of elastic collision between a pair of interacting particles has numerous applications, in particular to the transport properties of gases, plasmas, simulation of rarefied gas flows, etc. When classical description is sufficient and the interaction potential $U(r)$ is isotropic, the problem is equivalent to the
scattering of a single particle of reduced mass $\mu$ in a central force field. The scattering angle $\chi$ depends on the impact parameter $\rho$ and the kinetic energy of colliding particles, $\tfrac{1}{2}\mu v^2$, as~\cite{Landau2}
\begin{equation}\label{scattering_angle}
\chi(\rho)=|\pi-2\varphi(\rho)|, ~ \varphi(\rho)=\rho\int_{r_0}^\infty \frac{dr}{r^2\sqrt{1-U_{\rm eff}(r,\rho)}},
\end{equation}
where $U_{\rm eff}$ is the reduced effective potential energy
\begin{equation}
U_{\rm eff}(r,\rho)={\rho^2}/{r^2}+{2U(r)}/{\mu v^2}. \label{Ueff}
\end{equation}
Integration in (\ref{scattering_angle}) is performed from the distance of the
closest approach, $r_0(\rho)$ -- the largest root of the equation
\begin{equation}
U_{\rm eff}(r,\rho)=1. \label{closestapproach}
\end{equation}
Using Eqs.~(\ref{scattering_angle})-(\ref{closestapproach}), the dependence $\chi(\rho)$
can be calculated (at least numerically) for an arbitrary pair interaction potential $U(r)$.
The transport cross sections can then be obtained as proper integrals over the impact parameters.
The two key quantities associated with the transport coefficients (in the binary collision approximation) are the diffusions (momentum transfer) cross section
\begin{equation}
\sigma_{\rm D}=2\pi\int_0^\infty [1-\cos\chi(\rho)]\rho d\rho
\label{crossectionD}
\end{equation}
and the viscosity cross section
\begin{equation}
\sigma_{\eta }=2\pi\int_0^\infty [1-\cos^2\chi(\rho)]\rho d\rho.
\label{crossectionV}
\end{equation}
These cross sections are velocity-dependent. Their names come from the fact that the diffusion and viscosity coefficients can be obtained by properly integrating these cross sections over the distribution of velocities.

The diffusion and viscosity cross sections have been calculated for many potentials that are used to model interactions in real systems. Perhaps, the most studied case is the conventional $12-6$ Lennard-Jones (LJ) potential~\cite{LJ} of the form
\begin{equation}
U(r)=4\epsilon\left[(d/r)^{12}-(d/r)^{6} \right],
\end{equation}
where $r$ is the separation between the particles, $\epsilon$ and $d$ are characteristic energy and length scales. A number of values of the cross sections along with the resulting transport integrals were tabulated (see e.g. Refs.~\cite{Hirschfelder,Smith,HirschBook}), in particular for the high-energy regime.

Accurate analytical expressions for the transport cross sections can be of considerable value when searching for simplified collision models to enhance the accuracy and efficiency of Monte Carlo simulations of rarefied gas flows. In particular, this includes the so-called variable hard-sphere and variable soft-sphere models as well as their generalizations and modifications~\cite{Bird,Hassan,Kunc,Koura,Matsumoto,Fan}. They can be also of practical use in other situations, an example will be given in this work.

The purpose of this paper is to introduce simple and accurate analytical expressions for the diffusion and viscosity cross sections for the LJ potential, which are convenient for practical implementation. In order to do this, the cross sections are re-evaluated numerically in a very wide range of energies. The limits of high and low energy scattering are then analyzed to identify the corresponding asymptotic behavior. It is then suggested how to modify the asymptotes in order their combination fits the numerical results with excellent accuracy (deviations do not exceed  $\pm 1\%$). Examples of using the proposed expressions are given towards the end of the paper.

\section{Numerical results}

\begin{figure}
\includegraphics[width=8.5cm]{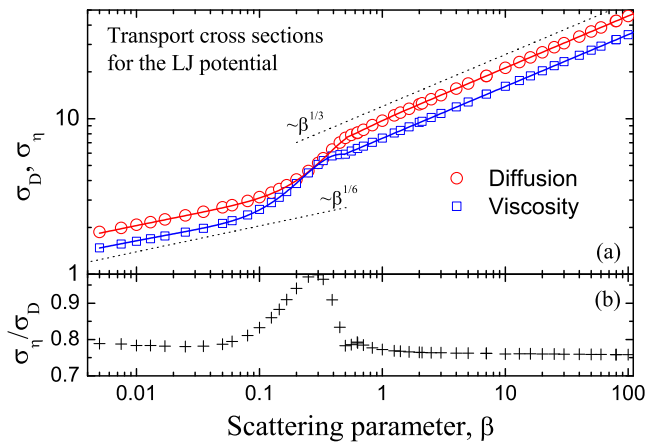}
\caption{(a) Diffusion and viscosity cross sections ($\sigma_{\rm D}$ and $\sigma_{\eta}$, respectively) as functions of the scattering parameter $\beta$. Symbols correspond to numerical results, curves are the fits. (b) Ratio of the viscosity-to-diffusion cross sections ($\sigma_{\eta}/\sigma_{\rm D}$) as a function of $\beta$.
} \label{f1}
\end{figure}

A portion of the obtained numerical results is shown in Fig.~\ref{f1}(a). Symbols correspond to the diffusion and viscosity cross sections (here and throughout the paper, the cross sections are in units of $d^2$), plotted versus the scattering parameter $\beta$, defined as
\begin{equation}\label{beta}
\beta(v)\equiv\epsilon/\mu v^2.
\end{equation}
From the definition (\ref{beta}) it is clear that $\beta$ measures the relative energy of colliding particles. The high-energy collision occurs when $\beta\ll 1$, while the low-energy collision takes place when $\beta\gg 1$. Note that the reduced transport cross sections depend only on $\beta$~\cite{Khrapak2014}.

Figure \ref{f1}(b) shows the ratio $\sigma_{\eta}/\sigma_{\rm D}$ as a function  of $\beta$. This ratio is almost constant for $\beta\lesssim 0.1$ and $\beta\gtrsim 1$, but exhibits a pronounced non-monotonic behavior in the transitional regime.

The present numerical results have been compared with those available in the literature. For example, Table~\ref{Tab2} lists the values of the diffusion cross section obtained in this work and those from Ref.~\cite{Smith}. The maximum deviations do not exceed $0.2\%$. The comparison has also been performed with the values of diffusion and viscosity cross sections tabulated in the classical paper by Hirschfeleder {\it et al}.~\cite{Hirschfelder}. The agreement is normally better than to within $0.1\%$ in the high-energy regime, but becomes worse when $\beta$ approaches unity. The maximum relative deviation can be as high as $\simeq 5\%$ for the diffusion cross section at $\beta=2.5$. This observation correlates well with the earlier conclusion from Ref.~\cite{Storck} that the results of Ref.~\cite{Hirschfelder} are not very accurate in the regime of moderate energies (see Figures 10 and 11 from Ref.~\cite{Storck}). Overall, the obtained numerical results are expected to have an accuracy of $\sim {\mathcal O}(0.1\%)$, which is more than sufficient for the main purpose of this study.

\begin{table}[t!]
\caption{\label{Tab2} Diffusion (momentum transfer) cross section $\sigma_{\rm D}$  obtained in the present work and in Ref.~\cite{Smith} for several values of the scattering parameter $\beta$.}
\begin{ruledtabular}
\begin{tabular}{lll}
$\beta$ & Present work & Ref.~\cite{Smith}  \\ \hline
79.365 & 42.41 & 42.34   \\
7.042 & 18.87  & 18.89   \\
2.099 & 12.58  & 12.58    \\
0.626 & 8.147  & 8.166    \\
0.391 & 6.331  & 6.337    \\
\end{tabular}
\end{ruledtabular}
\end{table}

\section{Low- and High-energy limits}

\begin{figure}
\includegraphics[width=8cm]{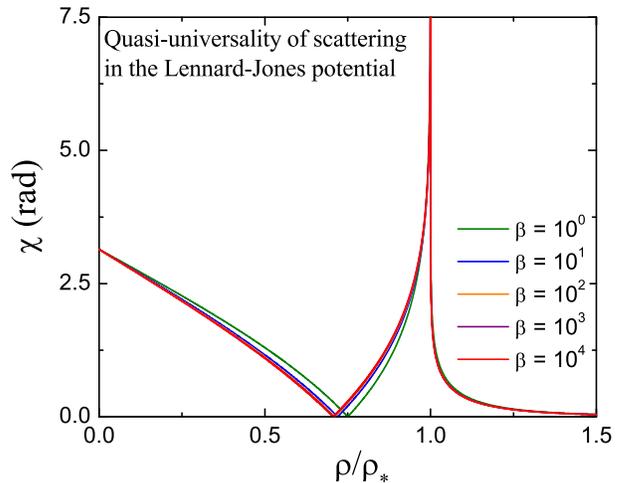}
\caption{Scattering angle $\chi$ versus the reduced impact parameter $\rho/\rho_*$, where $\rho_*$ is the critical impact parameter corresponding to orbiting trajectories. Curves are plotted for five different values of the scattering parameter $\beta$. Except for the lowest value $\beta=1$, all other curves are practically indistinguishable from each other.
} \label{f2}
\end{figure}

The detailed investigation of the low-energy (high-$\beta$) limit has been reported recently~\cite{Khrapak2014}. One of the most important results from this study is that the scattering angle $\chi$ becomes a quasi-universal function of the suitably reduced impact parameter $\rho$. The proper normalization is the critical impact parameter $\rho_*$, corresponding to a barrier in the effective potential energy, which results in orbiting trajectories and divergence of the scattering angle. Although the LJ potential is not the unique interaction potential exhibiting such a property (e.g. Yukawa and exponential interactions also demonstrate quasi-universality~\cite{Khrapak2014,id2,id3}), the universality of this kind is most pronounced in the LJ case. The illustration of this quasi-universality is given in Fig.~\ref{f2}.

The quasi-universality implies that the transport cross sections scale as
\begin{equation}\label{LE}
\sigma_{{\rm D},\eta}^{\rm LE}\simeq {\mathcal A}_{{\rm D},\eta}\pi\rho_*^2,
\end{equation}
where ${\mathcal A}_{\rm D}$ and  ${\mathcal A}_{\eta}$ are very weak functions of $\beta$, which can be easily evaluated numerically. For most practical purposes they can be assumed constant
${\mathcal A}_{\rm D}\simeq 0.83$ and ${\mathcal A}_{\eta}\simeq 0.63$. There is a simple analytical relation between the critical impact parameter $\rho_*$ and the scattering parameter $\beta$.
In the limit of large $\beta$ it yields the scaling $\rho_*\simeq 1.94\beta^{1/6}$~\cite{Khrapak2014}. The resulting dependence $\sigma_{{\rm D},\eta}^{\rm LE}\propto \beta^{1/3}$ in the low-energy regime is indicated in Fig.~\ref{f1}(a).

In the opposite high-energy limit (low-$\beta$ limit) the cross sections are mainly determined by the short-range $r^{-12}$ repulsion. The proper interaction length scale is therefore
$R\simeq (8\varepsilon/mv^2)^{1/12}\propto \beta^{1/12}$. The scaling of the cross sections is thus
\begin{equation}\label{HE}
\sigma_{{\rm D},\eta}^{\rm HE}\simeq {\mathcal B}_{{\rm D},\eta} \beta^{1/6}.
\end{equation}
The coefficients of proportionality (which are different for diffusion and viscosity) are {\it non-monotonous}, but rather weak function of $\beta$. The resulting scaling $\sigma_{{\rm D},\eta}^{\rm HE}\propto \beta^{1/6}$ in the high-energy regime is also indicated in Fig.~\ref{f1}(a).

Figure~\ref{f1}(a)  shows that the low-energy asymptote of Eq.~(\ref{LE}) describes well the numerical results for $\beta\gtrsim 1$, while the high-energy asymptote of Eq.~(\ref{HE}) is well applicable at $\beta\lesssim 0.1$. In the next section fits are proposed, which take into account the discussed asymptotic behavior and describe accurately the transition between these two limits.

\section{Accurate fits}

In constructing the fits, one obvious possibility is to just take a sum of low- and high-energy asymptotes in the form
\begin{equation}\label{sum1}
\sigma_{{\rm D},\eta}=\sigma_{{\rm D},\eta}^{\rm LE}+\sigma_{{\rm D},\eta}^{\rm HE}.
\end{equation}
This would essentially resemble the energy dependence of the transport cross sections in the generalized hard-sphere model of Ref.~\cite{Hassan} and generalized soft-sphere model of Ref.~\cite{Fan}, where analogous two-term formulas have been used. However, although equation (\ref{sum1}) is essentially exact in the respective limits, it is very crude in the wide transitional regime, so one needs to look for another option.

\begin{figure}[t!]
\includegraphics[width=8cm]{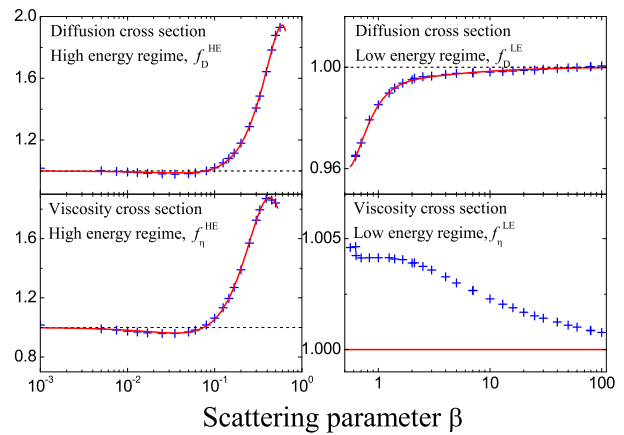}
\caption{Correction functions used to produce accurate fits for the diffusion and scattering cross sections. Symbols correspond to the numerical results, curves are the fits using Eqs.~(\ref{fit1}) and (\ref{fit2}) with the coefficients summarized in Table~\ref{Tab1}. Taking into account the required accuracy, we just take $f_{\eta}^{\rm LE}=1$ for the viscosity cross section in the low-energy regime.}
\label{f3}
\end{figure}

The strategy to be implemented here is similar to that applied recently in the case of the attractive Yukawa potential~\cite{Yukawa}. We consider two regimes, one with $\beta\lesssim 0.5$ and the other with $\beta \gtrsim 0.5$. In the first regime we take the high energy asymptotes $\sigma_{{\rm D},\eta}^{\rm HE}$ as the basis, and introduce the correction functions $f_{{\rm D},\eta}^{\rm HE}$ to improve the agreement with numerical data in this regime. Similarly, in the second regime, we look for a correction functions $f_{{\rm D},\eta}^{\rm LE}$ which allow to extend the applicability of the $\sigma_{{\rm D},\eta}^{\rm LE}$ asymptote into the regime of moderate $\beta$.
The two expression are then matched around $\beta\simeq 0.5$. The correction functions are assumed to have the following simple form:
\begin{equation}\label{fit1}
f_{{\rm D},\eta}^{\rm LE} =1+ \sum_{i=i}^{4}c_i\beta^{-i}
\end{equation}
in the low-energy regime, and
\begin{equation}\label{fit2}
f_{{\rm D},\eta}^{\rm HE} = 1+\sum_{i=1}^{4}c_i\beta^i
\end{equation}
in the high-energy regime. These forms ensure that no unphysical behavior occurs in the respective limits.
The coefficients for these correction functions are summarized in Table~\ref{Tab1}. Figure~\ref{f3} shows how these functions agree with the numerical results.
Note that to within the required accuracy the correction function for the viscosity cross sections in the low-energy regime can be just set to unity,
$f_{\eta}^{\rm LE}=1$.

The resulting fits for the diffusion and viscosity cross sections for the LJ potential are
\begin{equation}\label{Diffusion}
\sigma_{\rm D}(\beta) = \begin{cases} 4.507\beta^{1/6}f_{\rm D}^{\rm HE}, & \beta< 0.506 \\ 9.866\beta^{1/3}f_{\rm D}^{\rm LE}, & \beta>0.506 \end{cases}
\end{equation}
and
\begin{equation}\label{viscosity}
\sigma_{\eta}(\beta) = \begin{cases} 3.599\beta^{1/6}f_{\eta}^{\rm HE}, & \beta< 0.491 \\ 7.480\beta^{1/3}f_{\eta}^{\rm LE}, & \beta>0.491 \end{cases}
\end{equation}
These expressions constitute the main result of this paper. They are shown by solid curves in Fig.~\ref{f1}(a), the agreement with numerical results is excellent. It has been verified that for $\beta\gtrsim 10^{-3}$ the deviations between these simple fits and numerical results do not exceed $\pm 1\%$, and for most of the data points are considerably smaller than that, especially in the low-energy regime~\cite{Com1}. The collision integrals result from the integration of these cross sections with the velocity distribution function, and thus should be even more accurate.

\begin{table}
\caption{\label{Tab1} Fitting parameters entering equations (\ref{fit1}) and (\ref{fit2}).}
\begin{ruledtabular}
\begin{tabular}{llllll}
Function &  $c_1$ & $c_2$ & $c_3$ & $c_4$   \\ \hline
$f_{\rm D}^{\rm HE}$  & -0.692  & 9.594  & -8.284  & -2.355   \\
$f_{\rm D}^{\rm LE}$  & -0.019  & 0.038  & -0.049  &  0.015   \\
$f_{\eta}^{\rm HE}$   & -2.229  & 35.967 & -86.490 &  60.335  \\

\end{tabular}
\end{ruledtabular}
\end{table}

\section{Example I: Self-diffusion and viscosity of Argon}

The Lennard-Jones pair potential is known to be a simple and reasonable approximation (although not ideal) of actual interactions in non-polar substances. In this section the self-diffusion ($D_{\rm s}$) and viscosity ($\eta$) coefficients of argon gas are calculated from the transport cross sections derived above and compared with the data available in the literature. In particular, the reference data summarized in Ref.~\cite{Kerstin} are used for comparison with the actual transport coefficients. The numbers from Table IV of Ref.~\cite{Fan} are used to compare our results for the self-diffusion coefficient with experiment and predictions of some other models. For consistency in this latter comparison, the same LJ parameters for argon as in Ref.~\cite{Fan} are adopted: $\epsilon = 124$ K, $d= 3.418$ {\AA}.

The self-diffusion coefficient can be expressed as
\begin{equation}\label{Ds}
D_{\rm s}= \frac{3\sqrt{\pi}}{8}\frac{T^{3/2}}{p m^{1/2}d^2\Omega_{\rm D}},
\end{equation}
where $\Omega_{\rm D}$ is the reduced diffusion integral
\begin{equation}\label{d_int}
\Omega_{\rm D}=\tfrac{1}{2}\int_0^{\infty}x^2e^{-x}\sigma_{\rm D}(x)dx.
\end{equation}
Here $p$ is the pressure, $x=\mu v^2/2T=(2T_* \beta)^{-1}$ is the reduced energy, $T_*=T/\epsilon$ is the reduced temperature, and it is taken into account that $\mu=m/2$ for identical particles.
The diffusion integral (as all other transport integrals) is a function of $T_*$. It can be easily calculated by substituting $\beta=(2T_*x)^{-1}$ into Eq. (\ref{Diffusion}) and integrating.
Similarly, the viscosity coefficient can be written as
\begin{equation}\label{visc}
\eta = \frac{5\sqrt{\pi}}{8}\frac{T^{1/2}m^{1/2}}{d^2\Omega_{\eta}},
\end{equation}
where $\Omega_{\eta}$ is the reduced viscosity integral
\begin{equation}\label{v_int}
\Omega_{\eta}=\tfrac{1}{2}\int_0^{\infty}x^3e^{-x}\sigma_{\eta}(x)dx,
\end{equation}
which can be evaluated using the fit (\ref{viscosity}). Equations (\ref{Ds}) and (\ref{visc}) correspond to the dominant (first-order) terms in the Chapman-Enskog approximation. Higher order corrections are neglected.

\begin{figure}[t!]
\includegraphics[width=8cm]{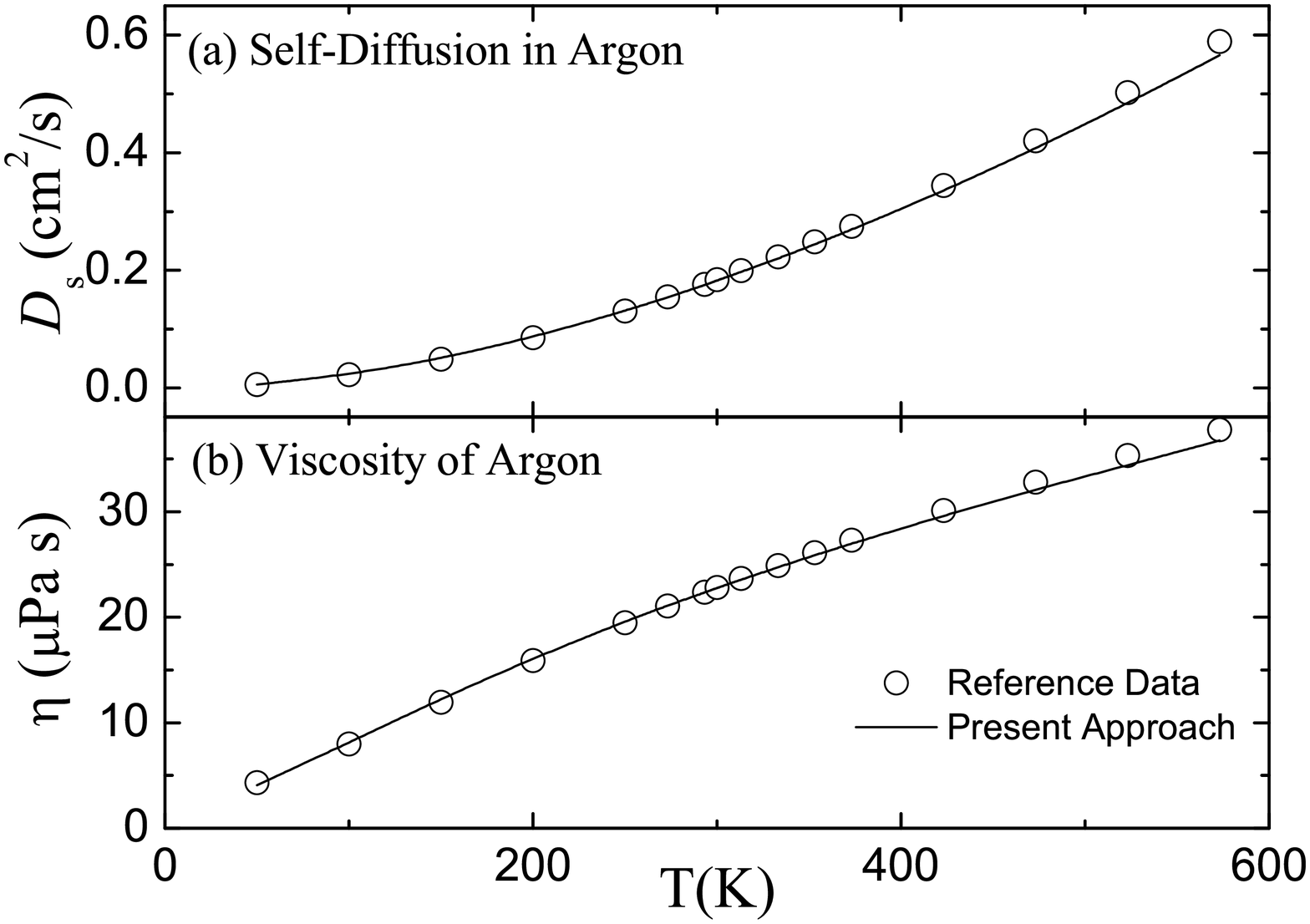}
\caption{(a) Self-diffusion coefficient in argon versus temperature at $p=1.013$ bar. (b) Viscosity of argon versus temperature. Symbols correspond to the data from Ref.~\cite{Kerstin}.  Solid curves are calculated using Eqs.~(\ref{Diffusion})-(\ref{v_int}).}
\label{f4}
\end{figure}

Figure~\ref{f4} shows the comparison between the calculation using Eqs. (\ref{Diffusion}) - (\ref{v_int}) and reference data for the self-diffusion and viscosity of argon from the Table 3 of Ref.~\cite{Kerstin}. The overall agreement is quite good. In the examined temperature range the maximum deviations ($\simeq 6\%$) are observed at the lowest temperature of $T=50$ K (in some contrast to the visual impression from  Fig.~\ref{f4}). This can be an indication that the quantum effects become important in this regime and the picture of classical scattering is insufficient~\cite{Imam}. At higher temperatures, there is a relatively wide range (up to $T\simeq 400$ K for diffusion and $T\simeq 500$ K for viscosity) where deviations are mostly within $\simeq 1 - 2\%$. At even higher temperatures, the present approach underestimate the coefficients of self-diffusion and viscosity by several percents. This is likely due to deviations in the actual interaction from the model $\propto r^{-12}$ repulsive LJ term.

In Ref.~\cite{Fan} the comparison between experimental results for the self-diffusion coefficient and predictions of several models used in Monte Carlo simulations of rarefied gas flows was made (see Table IV in ~\cite{Fan}). Among the models compared, the generalized soft sphere (GSS)~\cite{Fan} and the variable soft sphere (VSS)~\cite{Koura} models demonstrated better agreement with the experimental results. We reproduce the values for argon in Table~\ref{Tab3}, along with the present calculation based on Eqs.~(\ref{Diffusion}), (\ref{Ds}) and (\ref{d_int}).
For all three temperatures our calculation results are in better agreement with the experimental data.

\begin{table}[t!]
\caption{\label{Tab3} Self-diffusion coefficient (in cm$^2$/s) of argon gas at $p=1.013$ bar: Comparison between three theoretical approaches (Present work, GSS, and VSS) and experiment. The values corresponding to GSS, VSS, and experiment are taken from the Table IV of Ref.~\cite{Fan}.}
\begin{ruledtabular}
\begin{tabular}{lllll}
$T$ (K) & Present & GSS & VSS & Experiment  \\ \hline
77.7  & 0.0133 & 0.0130 & 0.0178 & 0.0134$\pm$0.0002   \\
273.2 & 0.154  & 0.161  & 0.173  & 0.156$\pm$0.002    \\
353.2 & 0.244  & 0.258  & 0.276  & 0.249$\pm$0.003    \\
\end{tabular}
\end{ruledtabular}
\end{table}

Situation becomes more complicated when deviations from the ideal gas behavior become significant. At present there is no unifying quantitative description of transport phenomena in condensed matter. Methods that have been used to describe transport phenomena in liquids include for instance the Enskog theory for hard sphere fluid and its modifications~\cite{Dymond,Miyazaki}, empirical fits based on the results from computer simulations~\cite{Levesque}, as well as various corresponding-states relationships. Among them the excess entropy scaling of the transport coefficients~\cite{Rosenfeld,Dzugutov} can be particularly mentioned since it applies to a wide class of simple fluids, with quite different interactions. We do not elaborate on this further since the main focus here is on the weakly coupled (gaseous) LJ systems. 

\section{Example II: Negative Thermophoresis}

As another simple example, let us apply the obtained results to the problem of thermophoresis.
Thermophoresis describes the phenomenon wherein small particles (or molecules), suspended in a gas or liquid where a temperature gradient exists (but macroscopic flows are absent), drift along this temperature gradient~\cite{Talbot,Duhr}. The drift is normally from the side of high to the side of low temperature, but in some parameter regimes it can reverse the direction, which is then called negative thermophoresis~\cite{Duhr,Wang}.

In the free molecular regime, the thermophoretic force responsible for the drift can be expressed via the momentum transfer cross section as~\cite{Wang,KhrapakPoP2013}
\begin{equation}\label{TP}
F_{\rm T}=\frac{16\kappa\nabla T}{15\sqrt{2\pi}v_{T}}\int_0^{\infty}x^2(\tfrac{5}{2}-x)e^{-x}\sigma_{\rm D}(x)dx,
\end{equation}
where $\kappa$ is the gas thermal conductivity, $T$ is the temperature, and $v_{T}=\sqrt{T/m}$ is the gas thermal velocity ($\mu\simeq m$ when the particle is much heavier than the atoms or molecules of the gas). The velocity dependence of the momentum transfer cross section is a key factor, which governs the direction of the force. For relatively big spherical particles, the gas-particle collisions can be approximated as a hard sphere scattering, with the constant momentum transfer cross section ($\sigma_{\rm D}={\rm const}$). Then Eq.~(\ref{TP}) reduces to the well-known Waldmann's expression~\cite{Waldmann} and the force pushes the particles towards low temperatures. On the other hand, when dealing with charged particles in a plasma environment -- the so-called complex (dusty) plasmas~\cite{FortovUFN,FortovPR,FortovBook}-- the thermophoretic force can be associated not only with the neutral component~\cite{Jellum,Rothermel}, but also with the electron and ion components. In this case, the momentum transfer is dominated by Coulomb scattering and the thermal force associated with electrons and ions will normally push the particle towards the region of higher temperatures~\cite{KhrapakPoP2013}, implying negative thermophoresis. The physical reason behind this phenomena is that the (Coulomb) scattering momentum transfer cross section quickly decreases with velocity ($\sigma_{\rm D}\propto v^{-4}$), so that the momentum transfer from the colder side turns out to be more effective.

One can easily derive simple general criterion for the negative thermophoresis. Writing $\sigma_{\rm D}\propto x^{-\nu}\propto v^{-2\nu}$, the integral in Eq.~(\ref{TP}) can be evaluated explicitly to give
\begin{equation}
\int_0^{\infty}x^2(\tfrac{5}{2}-x)e^{-x}\sigma_{\rm D}(x)dx\propto (\nu-\tfrac{1}{2})\Gamma(3-\nu).
\end{equation}
It changes the sign at $\nu=1/2$, corresponding to the switch between positive and negative thermophoresis.

As the particle size decreases from micron-range to molecular scale, the model of hard sphere scattering in gas-particle collisions becomes progressively less and less accurate~\cite{Wang,Zhang}.
Lennard-Jones-type potentials may better describe gas-particle interactions in this regime. Although, the conventional (12-6) LJ potentials is likely not the best model for these interactions, we adopt it here for illustrative purposes. The numerical results plotted in Fig.~\ref{f1}(a) indicate that $\sigma_{\rm D}\propto v^{-1/3}$ in the high energy regime and $\sigma_{\rm D}\propto v^{-2/3}$ in the low energy regime, and according to the criterion above the conventional (positive) thermophoresis takes place. In the transitional regime, however, the decay of the cross section with velocity is more steep and here the negative thermophoresis can be expected~\cite{Wang}. We substitute expression (\ref{Diffusion}) into the integral in (\ref{TP}), perform the integration, and identify the conditions for negative thermophorersis. This results in the range $0.417\lesssim T_*\lesssim 0.951$, where negative thermophoresis occurs. These upper and lower boundaries, which are expected to be accurate to within $\simeq 1\%$, slightly improve the previous estimate from Ref.~\cite{Wang}, where the interval $ 0.45 \lesssim T_*\lesssim 0.95$ has been reported.

\section{Conclusion}

To summarize, simple analytical fits for the diffusion and viscosity cross sections for elastic scattering in the Lennard-Jones potential are proposed. The accuracy of these fits is better than $1\%$, which is sufficient for most practical applications. Two relevant illustrations are provided. The strategy used in this work can be easily applied to other model potentials of the Lennard-Jones type.

\begin{acknowledgments}
This work was partly supported by the Russian Foundation for Basic Research, Project No. 13-02-01099.
\end{acknowledgments}


\begin{thebibliography}{99}
\bibitem{Com1} By construction, the fit (\ref{fit2}) is not expected to give ``exact'' results in the high-energy limit, $\beta\rightarrow 0$. In this limit, the diffusion and viscosity cross sections are underestimated by about $3\%$.
\end{thebibliography}
\end{document}